\documentclass[doublecol]{epl2}
\usepackage{graphicx}
\usepackage{dcolumn}
\usepackage{bm}
\usepackage{epsf}
\usepackage{subfigure}
\usepackage{epstopdf}
\usepackage{amsmath}
\usepackage{amssymb}

\newcommand{\mean}[1]{\langle #1 \rangle}

\begin{document}
\title{On the relation between event-based and time-based current statistics}
\shorttitle{Event-based and time-based current statistics}

\author{Massimiliano Esposito\inst{1,2} \and Katja Lindenberg\inst{2} \and Igor M. Sokolov\inst{3}}
\shortauthor{M. Esposito \etal}
\institute{
 \inst{1} Center for Nonlinear Phenomena and Complex Systems, Universit\'e Libre de Bruxelles, Code Postal 231, Campus Plaine, B-1050 Brussels, Belgium.\\
 \inst{2} Department of Chemistry and Biochemistry and BioCircuits Institute, University of California, San Diego, La Jolla, CA 92093-0340, USA.\\
 \inst{3} Institut f\"ur Physik, Humboldt-Universit\"at zu Berlin, Newtonstrasse 15, D-12489 Berlin,
Germany}



\abstract{
Current statistics can be calculated in various ways. Event-based approaches use the statistics of the number of events occuring during a given time. Time-based approaches use the statistics of the time needed to reach a given number of events. By analyzing normal as well as anomalous statistics of nonequilibrium currents through a two level system in contact with two different reservoirs, we investigate the conditions under which these different statistics do or do not yield identical predictions. We rely on the continuous time random walk formulation introduced in our earlier work [Phys. Rev. E {\bf 77}, 051119 (2008)].
}

\maketitle
\section{Introduction}

The study of currents is of fundamental importance in statistical mechanics.
Average energy and particle currents are central in determining whether a system
is in equilibrium. Indeed, at equilibrium the net probability current between
any pair of system states is zero, i.e., detailed balance between all pairs of
states is satisfied.  This of course means that all net currents through the
system are zero. External driving mechanisms bring a system out of equilibrium
and may lead to non-zero currents.  Examples of driving mechanisms include
contacts with reservoirs of different temperatures or chemical potentials, and
external forces.
The resulting departure from detailed balance is responsible for
an associated non-zero entropy production~\cite{Schnakenberg}.

The study of current statistics of systems out of equilibrium has become an
important tool to analyze small systems where fluctuations around average behavior
can be significant. In macroscopic systems only average currents are of interest
because fluctuations are too insignificant to be observable. However, in small
systems the fluctuations of currents around the average provide rich
information about the nonequilibrium dynamics of these systems, as evidenced by the
considerable recent interest in the field of photon and electron counting
statistics~\cite{Gardiner00,BarkaiRev,Orrit,Nazarov07,EspositoReview}.
Furthermore, current fluctuations have recently been connected to the
generalized version of the Second Law of thermodynamics via a variety
of fluctuation theorems~\cite{Kurchan98,Lebowitz99,Crooks,Seifert05,
AndrieuxGaspard07a,EspositoHarbola07PRE}. However, there is a fundamental aspect
of current statistics that has, to our knowledge, not yet been fully explored,
namely, the circumstances under which ergodic conditions are or are not satisfied.
Ergodicity involves the equivalence between ensemble-averaged and
time-averaged statistical properties at equilibrium.
Here we extend the notion of ergodicity to nonequilibrium currents by
exploring conditions under which event-based and time-based current
statistics are asymptotically equivalent.

It is of particular interest to address these questions in the presence of anomalous
statistics~\cite{KlafterRev} that are known to significantly affect
the ergodic properties of the system~\cite{Dahan03,Barkai05,Klafter,Barkai08}.
Effects of anomalous statistics were first observed in averaged
current measurements on macroscopic disordered semiconductors and
amorphous solids, and were described in terms of
continuous time random walks (CTRW)~\cite{Scher,ScherMontroll,ScherShlesinger}.
Nowadays, anomalous statistics can be measured by photon counting experiments
on single nanodevices such as blinking single quantum dots~\cite{Kuno}, single
nanocrystals~\cite{Dahan03}, and single molecules~\cite{KlafterExp}.
Since these experiments can give access to the full probability
distribution of the fluctuations it becomes appropriate to extend
the concept of ergodicity beyond the average to higher moments.

\section{Definitions and main results}

To describe the issue, we introduce a system in which certain elementary events
occur as time $t$ progresses, for instance, the net transfer of a
mass or of a charge in or out of a system.  The process is
stochastic, that is, there is a distribution of the number of events
occurring in a given time interval, or there is a distribution of times at which
one observes a given number of events. The current associated with such events
is usually defined as $I=k/t$ under the assumption that the number of events,
$k$, scales linearly with time. We will call this the ``normal'' scaling,
and generalize the definition of the current to recognize the possibility
that the number of events does not grow linearly with time,
\begin{equation}
I = \frac{k}{t^\alpha}.
\label{StochCurr}
\end{equation}
Normal behavior thus corresponds to $\alpha=1$, and we call other scaling
behaviors ``anomalous''. In particular, we focus on anomalous behavior with
sublinear scaling, $0 < \alpha <1$.
As we shall see, this scaling is consistent with the CTRW formalism and, more specifically,
with the subordination principle of so-called
anomalous statistical processes~\cite{Baeumer,Stanislavsky,Gorenflo}.

The scaling information is not sufficient to describe the full statistics
of the current. We need to be more precise about the nature of the distribution
that defines the behavior of the current and current moments.
The traditional approach to current statistics is to count the number of events
that occur during a given time interval. Mathematically, this means that one
treats $k$ as a random variable and $t$ as a parameter. The associated
probability distribution is ${\cal P}_t(k)$, the probability that $k$
events occur in time $t$.
The current moments according to this point of view are calculated as
\begin{equation}
\langle I^m\rangle_t \equiv \frac{\langle k^m\rangle_t}{t^{\alpha
m}}=\frac{1}{t^{\alpha m}} \sum_k k^m {\cal P}_t(k),
\label{momentEvents}
\end{equation}
where $m$ is a positive integer.
The average current is given by the first moment, $\mean{I}_t$.
The subscript serves as a reminder that time here is a parameter.

An alternative approach is to consider time to be the random variable and the number
of events the parameter. The associated probability density is ${\cal P}_k(t)$, the
probability that a time $t$ is needed to observe $k$ events. The current
moments according to this point of view are calculated as
\begin{equation}
\mean{J^m}_{k} \equiv \frac{k^m}{\mean{t^{\alpha m}}}_k
=\frac{k^m}{ \int dt \; t^{\alpha m}\; {\cal P}_k(t)}.
\label{momentTime}
\end{equation}
With this approach, we use a different symbol for the current ($J$) simply to
stress the difference.
The average current is now given by the first moment $\mean{J}_{k}$.
We stress that (\ref{momentEvents}) and (\ref{momentTime})
require different types of measurement of the current.
The first measures the current during a fixed time interval $t$ and the
second
measures it until a given number of events $k$ has occurred. The statistics is of
course obtained by repeating the experiment many times or by doing it simultaneously
on independent copies of the system.

At this point it might be tempting to say that the current statistics are
{\it ergodic} when asymptotically the statistics of events and the statistics
of times lead to identical results,
\begin{equation}
\lim_{t \to \infty} \frac{\langle k^m\rangle_t}{t^{\alpha m}} =
\lim_{k\to\infty}\frac{k^m}{\mean{t^{\alpha m}}}_k,
\label{ErgodicCond}
\end{equation}
that is,
\begin{equation}
 \lim_{t \to \infty} \mean{I^m}_t = \lim_{k\to\infty}\mean{J^m}_{k}.
\label{ErgodicCondprime}
\end{equation}
To our knowledge this concept of ergodicity would be new for two reasons.
First, it defines an equivalence of time and ensemble averages at the level
of the currents instead of ordinary system observables. The time average of
a current cannot be expressed in terms of the fraction of time that the system
spends in a given state along a trajectory, as do conventional approaches to
ergodicity (e.g., \cite{Barkai05}). Furthermore, currents are not uniquely
specified by the system states and require that the reservoirs responsible
for the transitions be specified as well.
Second, we have not restricted this statement to averages ($m=1$)
but have stated it for all moments ($m \geq 1$).

One of our two main results of this paper is the following:
\begin{list}{}
\item {}
The ergodic condition~(\ref{ErgodicCond}) \emph{is} satisfied for normal statistics,
$\alpha=1$, but \emph{fails} for anomalous statistics, $0 <\alpha <1$.\\
\end{list}

Our second goal in this work is to introduce current statistics for which an
ergodic condition \emph{is} satisfied for the entire range $0< \alpha \leq 1$.
This is achieved by dealing with the current even more
directly, so that time as a random variable
governed by the probability density ${\cal P}_k(t)$ enters through the \emph{inverse} moments,
\begin{equation}
 \mean{I^m}_k \equiv k^m \mean{t^{-\alpha m}}
 = k^m \int dt \; \frac{{\cal P}_k(t)}{t^{\alpha m}}.
\label{new}
\end{equation}
An alternative ergodic condition would then be the assertion that asymptotically the
statistics of events and the statistics of inverse times lead to identical results,
\begin{equation}
\lim_{t \to \infty} \frac{\langle k^m\rangle_t}{t^{\alpha m}} =
\lim_{k\to\infty}k^m\mean{t^{-\alpha m}}_k,
\label{ErgodicCond2}
\end{equation}
that is,
\begin{equation}
 \lim_{t \to \infty} \mean{I^m}_t = \lim_{k\to\infty} \mean{I^m}_{k}.
\label{ErgodicCondprime2}
\end{equation}

The second main result of this paper is the following:
\begin{list}{}
\item{}
The ergodic condition~(\ref{ErgodicCond2})
\emph{is} satisfied for all values $0<\alpha\leq 1$.\\
\end{list}

\begin{figure}[h]
\centering
\rotatebox{0}{\scalebox{0.65}{\includegraphics{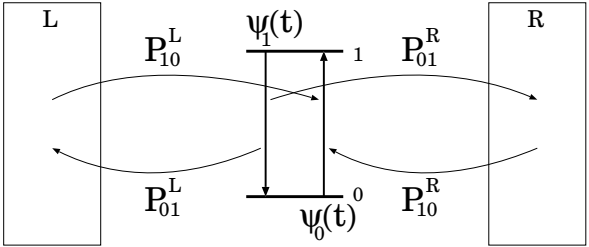}}}
\caption{Two-level system embedded between a left (L) and a right (R) particle
reservoir.}
\label{schema}
\end{figure}

While we could describe and justify these assertions in a very general way for currents
that scale as (\ref{StochCurr}), it
is more instructive to do so using a simple model that highlights the principal
issues. Our model system is sketched in Fig.~\ref{schema}, and falls within the
framework considered in Ref.~\cite{EspoLinden08}.

\section{The model}

The system consists of two levels, $i=0$ and $i=1$, connected to two
``particle'' reservoirs that we call left (L) and right (R). The nature
of the ``particles'' does not matter other than that the current is
associated with the flow of these particles between L and R through
the two-level system.
The nature of L and of R also does not matter other
than that L and R must of course be different in some way.
It is the asymmetry between the two that generates the
nonequilibrium constraint. We have chosen the simplest
possible system for our example, namely, one that allows the presence of
either no particles or one particle but no more than one particle.
Such a model can be viewed as a single-site asymmetric simple exclusion
process \cite{Derrida} and has been used, for example, to model electron
transport in a single-level quantum dot \cite{EspositoEPL}.

Transitions between the states of the system occur because of particle flow in
or out of the system to one or the other reservoir. We follow the
CTRW formulation in~\cite{EspoLinden08} (also called a
semi-Markov process, e.g., \cite{Maes,Andrieux}) to describe these transitions.
The transitions between the states are triggered by one or the
other of the reservoirs, and since these are distinct and correspond to
different mechanisms, we label the transitions with an index $\nu = $ L, R.
Suppose that the system arrives at state $i^\prime$ at a given time and that
its next transition is to state $i$ at a time $t$ later via mechanism $\nu$.
We assume the distribution $\psi_{i i^\prime}^\nu(t)$ for this
transition to be separable,
\begin{equation}
\psi_{i i^\prime}^\nu(t) = P_{i i^\prime}^\nu \psi_{i^\prime}(t).
\end{equation}
Here $P_{01}^\nu$ and $P_{10}^\nu$ are the transition probabilities downward
and upward, respectively, triggered by reservoir $\nu$, and the waiting time
distributions $\psi_0(t)$ for transitions
from $i=0$ to $i=1$ and $\psi_1(t)$ for transitions from $i=1$ to $i=0$ are
assumed to be independent of $\nu$.

To explore the consequences of anomalous statistics on various notions of
ergodicity, we choose waiting time distributions that decay asymptotically as
\begin{equation}
\psi_{i}(t) \sim (1/\tau_i)(t/\tau_i)^{-\alpha-1}.
\end{equation}
The times $\tau_0$ and $\tau_1$ are characteristic times (but not in general
first moments, which in fact diverge when $\alpha<1$).
As we shall see shortly, this scaling is consistent with the scaling
(\ref{StochCurr}) of the current.
The case $\alpha=1$ is
representative of waiting time distributions that decay at least as fast as
$1/t^2$, and these have a first moment $\tau_i$. In Laplace space,
to lowest order in $s$ the distributions behave as
\begin{eqnarray}
\tilde{\psi}_i(s) = 1 - (\tau_i s)^{\alpha}.
\label{WTDlaplace}
\end{eqnarray}
We note that a direct connection to thermodynamics can be made for $\alpha=1$
by, for instance, identifying the two reservoirs as heat baths of equal
temperatures $T$ but different chemical potentials $\mu_\nu$, and imposing
the conditions $(P_{10}^{\nu} \tau_1)/(P_{01}^{\nu} \tau_0)
=\exp{(\mu_{\nu}/k_B T)}$~\cite{EspoLinden08}.
A detailed discussion surrounding our choice of waiting time distribution can
be found in~\cite{Shlesinger}.

\section{Current statistics}

Having described our setup, we are now ready to calculate the various
statistical quantities needed to prove our statements about ergodicity.
First, consider event-based statistics. We define the number of events
$k_\nu$ to be the difference between the number of $0 \to 1$ and $1\to 0$
transitions (denoted $N_{01}$ and $N_{10}$) triggered by the $\nu$
reservoir by time $t$, i.e.,
\begin{eqnarray}
k_{\nu} = N_{01} P_{01}^{\nu} - N_{10} P_{10}^{\nu}.
\end{eqnarray}
It is always understood that a transition occurred at time $t=0$.
In Ref.~\cite{EspoLinden08} we used a generating function formalism to
obtain an analytic expression for the asymptotic form of the moments of
${\cal P}_{k_\nu}(t)$. Those results immediately lead to the result for the
left hand sides of Eq.~(\ref{ErgodicCond}) [or~(\ref{ErgodicCondprime})] and
Eq.~(\ref{ErgodicCond2}) [or~(\ref{ErgodicCondprime2})],
\begin{equation}
\lim_{t\to\infty} \mean{I_\nu^m}_t
= \lim_{t \to \infty} \frac{\langle k_\nu^m\rangle_t}{t^{\alpha m}}
= \frac{\Gamma(m+1)}{\Gamma(\alpha m+1)} a_\nu ^m,
\label{leftside}
\end{equation}
where
\begin{equation}
a_\nu \equiv \frac{P_{10}^\nu - P_{01}^\nu}{\tau_0^\alpha + \tau_1^\alpha}.
\label{definition}
\end{equation}

To calculate the right hand sides, we next consider the time-based statistics,
that is, the statistics of the times needed to reach a given number of events.
The probability $P_{k_\nu}(t)$ is the probability that $k_\nu$ events have
occurred by time $t$ and is simply a time convolution of $N_{10}$ factors
$\psi_0(t)$ and $N_{01}$ factors $\psi_1(t)$ in some order.
The order is intertwined with the events triggered by the other reservoir,
but does not matter when we go to Laplace space.
Recalling that convolutions in the time domain lead to products in Laplace space,
we get for the Laplace transform of $P_{k_\nu}$ the product
\begin{eqnarray}
\tilde{P}_{k_\nu}(s) = [\tilde{\psi}_0(s)]^{N_{10}}[\tilde{\psi}_1(s)]^{N_{01}}.
\label{detailed}
\end{eqnarray}
The numbers of events $N_{10}$ and $N_{01}$ are random variables.
However, this can be simplified at long times because whereas the total number
of transitions $2N=N_{01}+N_{10}$ is large, the difference is small, $\vert N_{01}-N_{10}
\vert \leq 1$, the constraint coming from the fact that no more than one
particle is allowed in the system. Therefore at long times we can set
$N_{01} \approx N_{10} \approx N$, so that
\begin{eqnarray}
\tilde{P}_{k_\nu}(s) = [\tilde{\psi}_0(s) \tilde{\psi}_1(s)]^{N}.
\end{eqnarray}
In the long time (small $s$) limit, Eq.~(\ref{WTDlaplace}) then leads to
\begin{eqnarray}
\tilde{P}_{k_{\nu}}(s) = [1-(\tau_0^{\alpha}+\tau_1^{\alpha}) s^{\alpha}]^{N},
\end{eqnarray}
which can be approximated by
\begin{eqnarray}
\tilde{P}_{k_{\nu}}(s) &\approx& \exp{[N \ln
\big(1-(\tau_0^{\alpha}+\tau_1^{\alpha})
s^{\alpha}\big)]} \nonumber \\
&\approx& \exp{[-N (\tau_0^{\alpha}+\tau_1^{\alpha}) s^{\alpha}]}
\nonumber\\
&=& \exp\left(- \frac{k_\nu}{a_\nu} s^\alpha\right).
\label{probexp}
\end{eqnarray}
In the last step we have set $k_\nu \approx N(P_{01}^\nu - P_{10}^\nu)$ and
have used Eq.~(\ref{definition}).

Equation~(\ref{probexp}) is the inverse Laplace transform of the one-sided
L\'evy distribution,
\begin{eqnarray}
{\cal P}_{k_{\nu}}(t) = \left(\frac{a_{\nu}}{k_{\nu}}\right)^{1/\alpha} {\cal
L}_{\alpha}
\bigg[ \left(\frac{a_{\nu}}{k_{\nu}}\right)^{1/\alpha}t \bigg].
\label{TimeDistrib}
\end{eqnarray}
The fractional moments of the L\'evy distribution for $0<\alpha \leq 1$
and $q<\alpha$ are~\cite{terdik,bardou}
\begin{equation}
\int du \; u^q\; {\cal L}_\alpha (u) = \frac{\Gamma(1-q/\alpha)}{\Gamma(1-q)},
\label{themoments1}
\end{equation}
from which it follows that
\begin{equation}
\mean{t^q}_{k_\nu} = \frac{\Gamma(1-q/\alpha)}{\Gamma(1-q)} \left(
\frac{k_\nu}{a_\nu}\right)^{q/\alpha}.
\label{themoments2}
\end{equation}
The higher positive moments ($q\geq \alpha$) are singular.
These moments allow us to calculate the right hand side of
of Eq.~(\ref{ErgodicCond}) [or Eq.~(\ref{ErgodicCondprime})].
In particular, the first
immediate conclusion is that for $0<\alpha <1$ the moments in the denominator
of the right hand side of Eq.~(\ref{ErgodicCond}) diverge.
This means that the right hand side of this presumptive
ergodicity condition vanishes, whereas the left hand side does not, cf.
Eq.~(\ref{leftside}).
Thus, the condition (\ref{ErgodicCond} [or (\ref{ErgodicCondprime})] indeed
fails for anomalous statistics.

On the other hand, when $\alpha=1$ one finds that for any value of $q$
\begin{equation}
\mean{t^q}_{k_\nu} = \left( \frac{k_\nu}{a_\nu}\right)^q,
\end{equation}
and hence we find for the right hand side of Eq.~(\ref{ErgodicCond}) that
\begin{equation}
\lim_{k_\nu \to \infty} \frac{k_\nu^m}{\mean{t^m}_{k_\nu}} = a_\nu^m.
\end{equation}
This is identical to the result obtained from Eq.~(\ref{leftside}) when
$\alpha=1$.  Thus, this condition is satisfied for normal
statistics.

Finally, we turn to the right hand side of Eq.~(\ref{ErgodicCond2}) or
(\ref{ErgodicCondprime2}).  The negative moments can be obtained from
Eqs.~(\ref{themoments1}) and (\ref{themoments2}) with $q<0$. We immediately
get
\begin{eqnarray}
\mean{t^{-\alpha m}}_{k_{\nu}} =  \frac{\Gamma(m+1)}{\Gamma(\alpha m+1)} \frac{a^{m}_{\nu}}{k^m_{\nu}} \label{LevyMoments}.
\end{eqnarray}
Using this result, we find that the time-based current statistics
leads to the moments
\begin{equation}
\lim_{k_\nu \to \infty} \mean{I_\nu^m}_{k_\nu}
= \lim_{k_\nu \to \infty} k_\nu^m \mean{t^{-\alpha m}}_{k_\nu}
=  \frac{\Gamma(m+1)}{\Gamma(\alpha m +1)} a_\nu^m.
\end{equation}
The last expression is identical to that occurring in Eq.~(\ref{leftside}).
This then proves the ergodic condition (\ref{ErgodicCond2}) or
(\ref{ErgodicCondprime2}) for all $0 < \alpha \leq 1$.

\section{Conclusions}

In this letter we have considered the current statistics of systems away
from equilibrium in which an external constraint produces a current. The
waiting time distribution between successive elementary processes that give rise
to the current decay as $t^{-\alpha-1}$ at long times, with $0<\alpha \leq 1$.
The scaling is ``normal" when $\alpha=1$ and ``anomalous" when $0 < \alpha <1$.
The elementary events may, for example, be a transfer of particles or of charge.
Specifically, we have addressed the question of the equivalence between different
ways of calculating the current statistics.

Two setups can be used to directly measure a (properly scaled) fluctuating current $I=k/t^\alpha$.
One can fix the time interval $t$ and measure the number $k$ of events occuring during this
time or, alternatively, one can fix the number of events $k$ and measure the time $t$ required
for this number of events to occur for the first time.
We have shown that in both of these setups, the average current as well as the
higher moments of the current are asymptotically the same, that is, $\lim_{t\to\infty}
\langle k^m \rangle/t^{\alpha m} = \lim_{k\to \infty} k_m \langle t^{-\alpha m}\rangle$ for all
$0<\alpha \leq 1$, cf. Eqs.~(\ref{momentEvents}), (\ref{new}), and (\ref{ErgodicCondprime2}).
A similar conclusion, but for a different model involving average particle
velocities in tilted periodic potentials, has been reached in~\cite{universal}.
When the current moments are calculated indirectly using the average time
statistics $k^m/\langle t^{\alpha m}\rangle$, cf. Eq.~(\ref{momentTime}),
difficulties arise in the case of anomalous scaling. This occurs because
this approach is in fact an incorrect way of calculating current moments
except for normal scaling, where all three calculations
of the current moments yield the same asymptotic results.

\acknowledgments

M. E. is supported by the FNRS Belgium (charg\'e de recherches) and by the
government of Luxembourg (Bourse de formation recherches). I. S. thanks the DFG
for support through the joint collaborative
program SFB 555.  K. L. gratefully
acknowledge the support of the US National Science Foundation through Grant No.
PHY-0855471.


\end{document}